\documentclass[english,reprint]{revtex4-2}

\usepackage{courier}

\usepackage[T1]{fontenc}
\usepackage[latin9]{inputenc}
\usepackage{geometry}
\usepackage{color}
\usepackage{babel}
\usepackage{multirow}
\usepackage{amsmath}
\usepackage{graphicx}
\usepackage[unicode=true,
 bookmarks=true,bookmarksnumbered=true,bookmarksopen=true,bookmarksopenlevel=1,
 breaklinks=true,pdfborder={0 0 1},backref=false,colorlinks=true]
 {hyperref}
\hypersetup{pdftitle={High temperature embedding sDFT},
 pdfauthor={Minh Nguyen},
 linkcolor=black,citecolor=black,urlcolor=blue,filecolor=blue,pdfpagelayout=OneColumn,pdfnewwindow=true,pdfstartview=XYZ,plainpages=false}

\makeatletter

\providecommand{\tabularnewline}{\\}

\usepackage{babel}
\usepackage{chemformula}
\makeatother

\begin{document}
\title{Tempering stochastic density functional theory}
\author{Minh Nguyen, Wenfei Li, Yangtao Li}
\affiliation{Department of Chemistry and Biochemistry, University of California
at Los Angeles, Los Angeles, California 90095, USA}
\author{Eran Rabani}
\affiliation{Department of Chemistry, University of California and Materials Sciences
Division, Lawrence Berkeley National Laboratory, Berkeley, California
94720, USA and The Raymond and Beverly Sackler Center of Computational
Molecular and Materials Science, Tel Aviv University, Tel Aviv 69978,
Israel}
\author{Roi Baer\textsuperscript{}}
\affiliation{Fritz Haber Center of Molecular Dynamics and Institute of Chemistry,
The Hebrew University of Jerusalem, Jerusalem, 91904 Israel}
\author{Daniel Neuhauser}
\thanks{Corresponding Author: Daniel Neuhauser, dxn@ucla.edu}
\affiliation{Department of Chemistry and Biochemistry, University of California
at Los Angeles, and California Nanoscience Institute, Los Angeles,
California 90095, USA }
\begin{abstract}
We introduce a tempering approach with stochastic density functional
theory (sDFT), labeled t-sDFT, which reduces the statistical errors
in the estimates of observable expectation values. This is achieved
by rewriting the electronic density as a sum of a ``warm'' component
complemented by ``colder'' correction(s). Since the warm component
is larger in magnitude but faster to evaluate, we use many more stochastic
orbitals for its evaluation than for the smaller-sized colder correction(s).
This results in a significant reduction in the statistical fluctuations
and systematic deviation compared to sDFT for the same computational
effort. We demonstrate the method's performance on large hydrogen-passivated
silicon nanocrystals, finding a reduction in the systematic deviation
in the energy by more than an order of magnitude, while the systematic
deviation in the forces is also quenched. Similarly, the statistical
fluctuations are reduced by factors of $\approx$4-5 for the total
energy and $\approx$1.5-2 for the forces on the atoms. Since the
embedding in t-sDFT is fully stochastic, it is possible to combine
t-sDFT with other variants of sDFT such as energy-window sDFT and
embedded-fragmented sDFT. 
\end{abstract}
\maketitle

\section{Introduction}

Kohn-Sham density functional theory (KS-DFT) is widely used for calculating
the properties of molecular and extended systems~\citep{jones2015density}.
In particular, the method is useful for determining the structure
based on the estimates it provides for the forces on the corresponding
nuclei~\citep{kolb2012molecular,selli2017modelling,freysoldt2014firstprinciples}.
However, applying KS-DFT for systems with hundreds or thousands of
atoms is challenging due to the high scaling of computational costs
with a system's size (potentially quadratic but eventually cubic for
large systems). Lower scaling implementations of the theory have been
developed for systems that have a density matrix that is fairly sparse
so that only a linear-scaling near-diagonal portion of the matrix
needs to be processed. Because of the restriction to near-localized
density matrices, the use of such methods is often limited to low-dimensional
structures~\citep{baer1997sparsity,scuseria1999linearscaling} or
systems with strictly localized electrons~\citep{ozaki2010efficient,romero-muniz2018highaccuracy}. 

In previous work we introduced stochastic density functional theory
(sDFT) ~\citep{baer2013selfaveraging} which avoids the costly diagonalization
step in KS-DFT without the need to make a locality assumption; instead,
the density matrix is approximated statistically. Specifically, the
density matrix in sDFT is viewed as a correlation function of stochastic
functions, each of which is, in essence, a random combination of the
occupied states. While the method scales linearly, the tradeoff is
the introduction of statistical uncertainties in the density and other
observables. The statistical errors can be reduced by using an embedded-fragmented
(ef-sDFT) technique \citep{neuhauser2014communication,chen2019overlapped,fabian2019stochastic}
which is based on dividing the system into fixed-size fragments and
expressing the total electron density, $n\left(\boldsymbol{r}\right)$,
as the sum of fragment densities plus a correction term which is evaluated
stochastically. This technique reduces the statistical fluctuations
in the estimates of the atomic forces and the energies~\citep{neuhauser2014communication,chen2019overlapped},
and the magnitude of this reduction is controlled by varying the size
of the fragments and the number of stochastic realizations. An additional
approach for mitigating the statistical errors is the energy-window
sDFT (ew-sDFT) scheme~\citep{chen2019energywindow} and its combination
with the embedded-fragmented technique~\citep{neuhauser2014communication,chen2021stochastic}.

Here we propose a tempering method, referred to as t-sDFT, as a complementary
technique for reducing the statistical noise. In t-sDFT, the density
for the desired temperature is calculated using a higher-temperature
reference density with smaller correction(s). This idea has been implemented
before within the energy renormalization group in the context of telescopically
expanding the Hamiltonian matrix in a series~\citep{baer1998energyrenormalizationgroup}.
In Sec.~\ref{sec:Methodology}, we describe the t-sDFT method and
in Sec. \ref{sec:Results} we benchmark and analyze its efficacy using
large hydrogenated silicon clusters. Sec.~\ref{sec:Summary-and-Discussion}
we discuss the conclusions and summarize.

\section{\label{sec:Methodology}Methodology}

\subsection{\label{subsec:Stochastic-Density-Functional}Stochastic Density Functional
Theory}

Our starting point is the following expression for the electron density,
$n\left(\boldsymbol{r}\right)$ (assuming a spin-unpolarized system)~\citep{baer2013selfaveraging}:

\begin{equation}
n\left(\boldsymbol{r}\right)=2\times\text{Tr}\left[\sqrt{\hat{\rho}_{\beta}}\left|\boldsymbol{r}\right\rangle \left\langle \boldsymbol{r}\right|\sqrt{\hat{\rho}_{\beta}}\right],\label{eq: original_density}
\end{equation}
where $\boldsymbol{r}$ is a point on a 3D grid that spans the space
containing the electron density of the system and has a volume element
$dV$. The operator
\begin{equation}
\hat{\rho}_{\beta}=f_{\beta\,\mu}\left(\hat{h}_{\text{}}\right)\label{eq:Filter}
\end{equation}
is a low band-pass Fermi-Dirac (FD) filter. Our main interest in this
paper is zero-temperature DFT; however, to efficiently represent the
density matrix, a smooth step function must be used, and the simplest
is a Fermi-Dirac distribution, $f_{\beta\,\mu}\left(\varepsilon\right)=\left(1+e^{\beta\left(\varepsilon-\mu\right)}\right)^{-1}$,
which blocks high energies ($\varepsilon>\mu+\beta$). Here, the chemical
potential, $\mu$, must be adjusted such that the integrated density
equals the number of electrons, $N_{e}$, 

\begin{equation}
\int n(r)dV=\sum_{\boldsymbol{r}}n\left(\boldsymbol{r}\right)dV=N_{e},\label{eq: integral of dens}
\end{equation}
while $\beta$ is the inverse temperature (note that the filter, $\hat{\rho}_{\beta}$,
depends on the chemical potential but to avoid a plethora of indices
we do not explicitly note this dependence below). In the low-temperature
limit ($\beta\varepsilon_{{\rm gap}}\gg1$, where $\varepsilon_{{\rm gap}}$
is the fundamental KS gap), $n\left(\boldsymbol{r}\right)$ indeed
converges to the ground state KS-density. Further, as our goal here
is not finite-temperature DFT, we use the usual zero-temperature DFT
exchange-correlation functionals.

In Eq.~\eqref{eq:Filter}, the Kohn-Sham Hamiltonian is:
\begin{equation}
\hat{h}_{{\rm }}=\hat{t}+\hat{v}\!\left[n\right]\!\left(\boldsymbol{r}\right)\label{eq:KS-H}
\end{equation}
where $\hat{t}$ is the kinetic energy operator and $\hat{v}\!\left[n\right]\!\!\left(\boldsymbol{r}\right)$
is the density-dependent KS potential that is composed of electron-nuclei,
Hartree, and exchange-correlation components. Eqs.~\eqref{eq: original_density}-\eqref{eq:KS-H}
must be solved simultaneously to yield the self-consistent electron
density.

In sDFT, we introduce a stochastic resolution of the identity~\citep{hutchinson1990astochastic}
which transforms the trace in Eq.~\eqref{eq: original_density} to
an expectation value~\citep{baer2013selfaveraging}

\begin{equation}
n\left(\boldsymbol{r}\right)=2\left\langle \left|\left\langle \boldsymbol{r}\left|\sqrt{\hat{\rho}_{\beta}}\right|\chi\right\rangle \right|^{2}\right\rangle _{\chi},\label{eq: stochastic density}
\end{equation}
where $\left|\chi\right\rangle $ is a stochastic orbital taking the
randomly signed values $\left\langle \boldsymbol{r}\left|\chi\right.\right>=\pm(dV)^{-1/2}$. 

To apply the filter, $\sqrt{\hat{\rho}_{\beta}}$, we use a Chebyshev
expansion of length $K$~\citep{kosloff1988timedependent}

\begin{equation}
\sqrt{\hat{\rho}_{\beta}}\left|\chi\right\rangle =\sum_{k=0}^{K}c_{k}\left(\beta,\mu\right)\left|\zeta^{\left(k\right)}\right\rangle ,\label{eq: chebyshev_exp}
\end{equation}
where $\left|\zeta^{\left(k\right)}\right\rangle =T_{k}(\hat{h}_{s})\left|\chi\right\rangle $
are defined by the Chebyshev polynomial recursion relations: $\left|\zeta^{\left(0\right)}\right\rangle =\left|\chi\right\rangle $,
$\left|\zeta^{\left(1\right)}\right\rangle =\left|\chi\right\rangle $
and $\left|\zeta^{\left(k+1\right)}\right\rangle =2\hat{h}_{s}\left|\zeta^{\left(k\right)}\right\rangle -\left|\zeta^{\left(k-1\right)}\right\rangle $.
Here, $\hat{h}_{s}=\left(\hat{h}-\bar{\varepsilon}\right)/\Delta\varepsilon$
is a normalized KS Hamiltonian where $\bar{\varepsilon}$ and $\Delta\varepsilon$
are chosen such that the spectrum of $\hat{h}_{s}$ lies within the
interval $\left[-1,1\right]$, $T_{k}(x)$ is the $k$'th Chebyshev
polynomial, and $c_{k}\left(\beta,\mu\right)$ are the Chebyshev expansion
coefficients of the filter $\sqrt{f_{\beta}\left(\varepsilon\right)}$\citep{kosloff1988timedependent}.
The expansion length $K$ terminates the series when $\left|c_{k>K}\right|$
is smaller than a predetermined cutoff value.

\begin{figure}
\begin{centering}
\includegraphics[width=1\columnwidth]{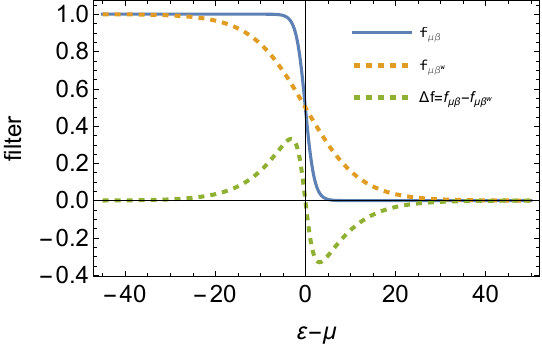}
\par\end{centering}
\caption{\label{fig:drho}The desired low-temperature filter, $f_{\beta\:\mu}\left(\varepsilon\right)$,
the high-temperature filter $f_{\beta^{w}\,\mu}\left(\varepsilon\right)$,
and the correction for $\beta=6\beta^{w}$.}
\end{figure}

In practice, the expected value appearing in Eq.~\eqref{eq: stochastic density}
is evaluated approximately by taking a finite sample of $N_{s}$ stochastic
orbitals: 
\begin{equation}
n\left(\boldsymbol{r}\right)\approx\frac{2}{N_{s}}\sum_{i=1}^{N_{s}}\left|\left\langle \boldsymbol{r}\left|\sqrt{\hat{\rho}_{\beta}}\right|\chi_{i}\right\rangle \right|^{2}.\label{eq:finiteSamplen(r)}
\end{equation}
Furthermore, to ensure $N_{e}=\sum_{\boldsymbol{r}}n\left(\boldsymbol{r}\right)dV$,
we tune the chemical potential $\mu$ to satisfy the relation:

\begin{equation}
N_{e}=2\sum_{k=0}^{K}b_{k}\left(\beta,\mu\right)M_{k},\label{eq: residues}
\end{equation}
where 
\begin{equation}
M_{k}=\frac{1}{N_{s}}\sum_{i=1}^{N_{s}}\left\langle \chi_{i}\left|\zeta_{i}^{\left(k\right)}\right.\right>\label{eq:Moments}
\end{equation}
are the stochastic estimates of the Chebyshev moments~\citep{sankey1994projected}
and $b_{k}\left(\beta,\mu\right)$ are the Chebyshev expansion coefficients
of the function $f_{\beta\,\mu}\left(\varepsilon\right)$ as opposed
to $c_{k}(\beta,\mu)$ which are the expansion coefficients of $\sqrt{f_{\beta\,\mu}(\varepsilon)}$. 

As a result of using stochastic orbitals, the sDFT density and associated
observables have two additional types of errors. One is the usual
stochastic fluctuations that scale as ${\rm O}(N_{s}^{-\frac{1}{2}})$,
but in addition, there is a systematic deviation which scales as ${\rm O}(N_{s}^{-1})$
that appears due to the nonlinear SCF procedure (the filtering operator
applied on each orbital depends on the density, which itself depends
on the set of filtered orbitals). 

Increasing the number of sampling orbitals, $N_{s},$ will decrease
both types of errors, at the cost of additional work. 

To measure the computational cost of a sDFT or t-sDFT calculation,
we use a numerical ``work'' quantity, $W$, which is the total number
of Hamiltonian operations performed per SCF cycle (i.e., action by
the Hamiltonian on a function), which for sDFT is approximately 
\[
W\simeq KN_{s}.
\]
In practice, the work needs to be multiplied by a factor of about
1.7 due to the need to determine $\mu$ based on Eq. (\ref{eq: residues}),
but since this factor is common to all our methods here we do not
include it.

\begin{table*}
\begin{centering}
\begin{tabular}{|c|c|c|c|c|c|c|c|c|c|}
\hline 
\multirow{1}{*}{System} & Band-gap & $\beta$ & \multicolumn{3}{c|}{Correction filter} & \multicolumn{3}{c|}{Warm filter} & \multirow{1}{*}{$W^{tot}$}\tabularnewline
\hline 
 & $({\rm eV)}$ & ${\rm (eV^{-1})}$ & $K$ & $N_{s}^{\Delta}$ & $W^{\Delta}=KN_{s}^{\Delta}$ & $K^{w}$ & $N_{s}^{w}$ & $W^{w}=K^{w}N_{s}^{w}$ & \tabularnewline
\hline 
\ch{Si35H36} & 3.4 & 1.83 & 2000 & 6 & $12,000$ & $K\times\beta^{w}/\beta$ & $24\times\beta/\beta^{w}$ & $48,000$ & $60,000$\tabularnewline
\hline 
\ch{Si87H76} & 2.5 & 2.94 & 3200 & $6$ & $19,200$ & $K\times\beta^{w}/\beta$ & $24\times\beta/\beta^{w}$ & $76,800$ & $96,000$\tabularnewline
\hline 
\ch{Si353H196} & 1.6 & 4.60 & 5000 & $6$ & $30,000$ & $K\times\beta^{w}/\beta$ & $24\times\beta/\beta^{w}$ & $120,000$ & $150,000$\tabularnewline
\hline 
\end{tabular}
\par\end{centering}
\caption{\label{tab:I}The Chebyshev expansion lengths, $K$ and $K^{w}$,
and the number of stochastic orbitals, $N_{s}^{\Delta}$ and $N_{s}^{w}$,
used in our simulations. We also show the required numerical work,
$W,$ defined as the number of Hamiltonian operations. Note that for
each system we increase the number of high-temperature orbitals ($N_{s}^{w}$)
with temperature (i.e., with increasing $\beta/\beta^{w})$ such that
the total work $W^{tot}$ is independent of $\beta/\beta^{w}$.}
\end{table*}

\subsection{\label{subsec:Tempering-Stochastic-Density}Tempering Stochastic
Density Functional Theory}

We now describe the tempering method, designed to reduce the statistical
errors in sDFT without increasing the overall computational effort.
Consider, the decomposition of the desired filter, $\hat{\rho}_{\beta}$
(see Eq.~\eqref{eq:Filter}), into a higher temperature filter, $\hat{\rho}_{\beta^{w}}$
($\beta^{w}<\beta$), with the correction term:

\begin{equation}
\Delta\hat{\rho}=\hat{\rho}_{\beta}-\hat{\rho}_{\beta^{w}},\label{eq: filter ht single corr}
\end{equation}
which is shown in Fig.~\eqref{fig:drho} for a typical case of $\beta/\beta^{w}=6$.
Note that (a) all values of $\Delta\hat{\rho}$ are much smaller than
unity and (b) the high-temperature filter, $\hat{\rho}_{\beta^{w}}$,
is smoother than the low-temperature one, so its Chebyshev expansion
is shorter (the Chebyshev expansion length of $\hat{\rho}_{\beta}$
is proportional to $\beta$~\citep{baer2013selfaveraging}). 

The electron density in Eq.~(\ref{eq: stochastic density}) can therefore
be written as $n\left(\boldsymbol{r}\right)=n_{\beta^{w}}\left(\boldsymbol{r}\right)+\Delta n\left(\boldsymbol{r}\right)$,
where the two terms are evaluated separately using two distinct independent
sets of stochastic orbitals: $\chi_{i}^{w}$, $i=1,\dots,N_{s}^{w}$
for the warmer density,
\begin{equation}
n_{\beta^{w}}\left(\boldsymbol{r}\right)=\frac{2}{N_{s}^{w}}\sum_{i=1}^{N_{s}^{w}}\left|\left\langle \boldsymbol{r}\left|\sqrt{\hat{\rho}_{\beta^{w}}}\right|\chi_{i}^{w}\right\rangle \right|^{2}\label{eq:warm_dens}
\end{equation}
and $\chi_{i}^{\Delta}$, $i=1,\dots,N_{s}^{\Delta}$ for the correction
term,
\begin{equation}
\Delta n\left(\boldsymbol{r}\right)=\frac{2}{N_{s}^{\Delta}}\sum_{i=1}^{N_{s}^{\Delta}}\left(\left|\left\langle \boldsymbol{r}\left|\sqrt{\hat{\rho}_{\beta}}\right|\chi_{i}^{\Delta}\right\rangle \right|^{2}-\left|\left\langle \boldsymbol{r}\left|\sqrt{\hat{\rho}_{\beta^{w}}}\right|\chi_{i}^{\Delta}\right\rangle \right|^{2}\right).\label{eq:Deltanr}
\end{equation}

\begin{figure}
\begin{centering}
\includegraphics{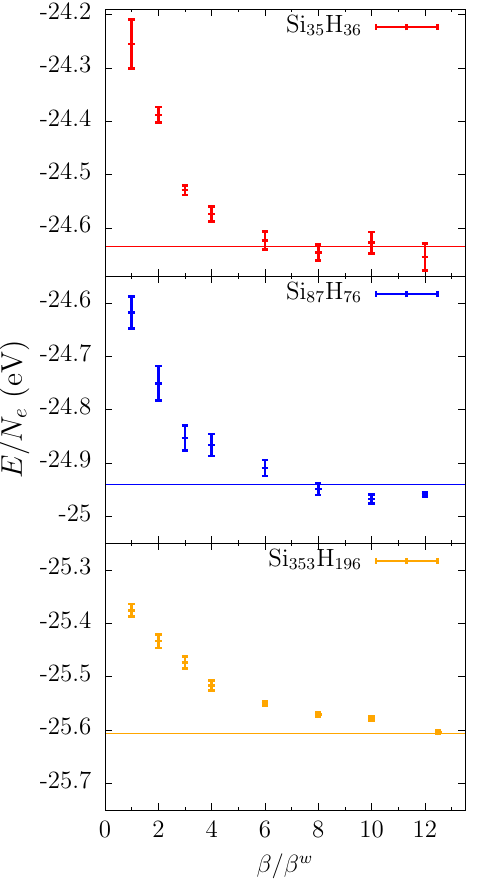}
\par\end{centering}
\caption{\label{fig:EperN}The energy per electron (in eV) as a function of
$\beta/\beta^{w}$ for three cluster sizes, based on $N_{{\rm ind}}=10$
independent runs. Also included is the deterministic value in each
system (horizontal line). The numerical work $W$, i.e., the number
of Hamiltonian operations, is independent of $\beta/\beta^{w}$ (see
Table~\ref{tab:I} for details). The leftmost point in each graph,
$\beta/\beta^{w}=1,$ corresponds to sDFT (no tempering). Since the
number of orbitals used is very small ($N_{s}=30$ for sDFT), these
sDFT results show marked systematic deviation (i.e., deviation of
the average energy from the deterministic value) and fluctuation errors.
Both types of statistical errors decrease markedly in t-sDFT, especially
when $\beta/\beta^{w}\sim7-10,$ due to the much larger number of
stochastic orbitals used in the main (warm) density part.}
\end{figure}

As demonstrated in Fig.~\ref{fig:drho}, the correction density,
$\Delta n\left(\boldsymbol{r}\right)$, is much smaller than the warm
density, $n_{\beta^{w}}\left(\boldsymbol{r}\right)$, which is similar
in overall magnitude to the total density. This gives the key for
the efficiency of the tempering approach as compared to the original
sDFT calculation. Specifically, compared to an sDFT calculation with
polynomial expansion length $K$ and $N_{s}$ stochastic orbitals
and aiming for the same overall work as in sDFT, we get that: 
\begin{itemize}
\item The computational work required to calculate the warm density is $W^{w}\simeq K^{w}N_{s}^{w}.$
Since the Chebyshev expansion lengths are proportional to $\beta$,
the warmer temperature density $n_{\beta^{w}}\left(\boldsymbol{r}\right)$
requires a much \emph{shorter }Chebyshev expansion length than the
original sDFT density ($K^{w}=\frac{\beta^{w}}{\beta}K$), so many
more stochastic orbitals can be used to evaluate it for the same overall
computational cost (i.e.$,N_{s}^{\Delta}\gg N_{s}$). 
\item The computational work for correction term, $\Delta n(\boldsymbol{r})$,
is $W^{\Delta}\simeq KN_{s}^{\Delta}$ as both terms in the RHS of
Eq. (\ref{eq:Deltanr}) use the same set of $\left|\zeta_{i}^{\Delta,(k)}\right\rangle =T_{k}(\hat{h}_{s})\left|\chi_{i}^{\Delta}\right\rangle $
(the two terms differ in their expansion coefficients). Since the
numerical magnitude of the correction term is much smaller than that
of the overall density, its standard deviation is correspondingly
much smaller. Therefore the numerical effort (i.e., the number of
samples required for a given accuracy), which is proportional to the
squared standard deviation, is much smaller for the correction term,
so it is sufficient to use fewer stochastic orbitals ($N_{s}^{\Delta}\ll N_{s}$)
to achieve a similar statistical error.
\end{itemize}
Here, $K^{w}$ and $K$ are respectively the Chebyshev expansion lengths
for the warm reference density and the correction term, with $\frac{K^{w}}{K}=\frac{\beta^{w}}{\beta}\ll1$
(note that since the correction term involves the original low-temperature
density, the number of Chebyshev terms it requires, $K$, is the same
as in the original sDFT). Overall, the partitioning of the filter
into a larger component at a higher temperature with a shorter Chebyshev
expansion, and a smaller correction term, offers an additional knob
to control the statistical error by using $N_{s}^{w}\gg N_{s}$ without
increasing the overall computational effort.

The use of tempering modifies how the chemical potential is calculated.
Instead of fulfilling the single-sum sDFT constraint on the residues
(Eq. (\ref{eq: residues})), the chemical potential is adjusted to
satisfy the following relation which consists of two summation terms
that each has its own Chebyshev expansion: 
\begin{align}
N_{e} & =2\sum_{k=0}^{K^{w}}b_{k}\left(\beta^{w},\mu\right)M_{k}^{w}\nonumber \\
 & \,\,\,\,\,\,+2\sum_{k=0}^{K}\left[b_{k}\left(\beta,\mu\right)-b_{k}\left(\beta^{w},\mu\right)\right]M_{k}^{\Delta}.\label{eq:Ne_t-sDFT}
\end{align}
The corresponding Chebyshev moments, defined in analogy to Eq.~\eqref{eq:Moments},
are: 
\begin{multline}
M_{k}^{w}=\frac{1}{N_{s}^{w}}\sum_{i=1}^{N_{s}^{w}}\left\langle \chi_{i}^{w}\left|\zeta_{i}^{w,(k)}\right.\right>,\,\,\,\,\,\,\,\,\,\\
M_{k}^{\Delta}=\frac{1}{N_{s}^{\Delta}}\sum_{i=1}^{N_{s}^{\Delta}}\left\langle \chi_{i}^{\Delta}\left|\zeta_{i}^{\Delta,(k)}\right.\right>.\label{eq:Moments_tsdft}
\end{multline}
In practice, finding the chemical potential in both sDFT and t-sDFT
involves a straightforward single-variable search. In t-sDFT, the
chemical potential (Eq. (\ref{eq:Ne_t-sDFT})) is not strictly guaranteed
to be monotonic with the number of electrons, since it is a summation
of two terms, one of which is not necessarily positive; but in practice,
we find that it is always monotonic so the determination of $\mu$
from the residues is instantaneous.

\section{\label{sec:Results}Results}

We studied three hydrogen-terminated silicon nanocrystals of different
sizes, \ch{Si35H36}, \ch{Si87H76}, and \ch{Si353H196}. Note
that such nanocrystals are a convenient test ground for stochastic
methods since they have small band gaps which decrease with increasing
system size. As such they place a more stringent test on the method
than clusters of molecules with large gaps. Metals, in contrast, require
much smaller temperatures and are therefore not ideal for stochastic
applications.

An LDA functional \citep{perdew1992accurate} was applied with norm-conserving
pseudopotentials \citep{troullier1991efficient}  using the Kleinman-Bylander
form~\citep{kleinman1982efficacious}, and we used the Martyna-Tuckerman
reciprocal-space method for treating long-range interactions~\citep{martyna1999areciprocal}.
The grid spacing was $0.55a_{0}$, and the energy cutoff was $15{\rm \:Hartree}$
for all systems. To gather sufficient statistics, $N_{{\rm ind}}=10$
independent runs with different stochastic numbers were used for each
of the calculations below. 

\begin{figure}
\begin{centering}
\includegraphics[width=1.03\columnwidth]{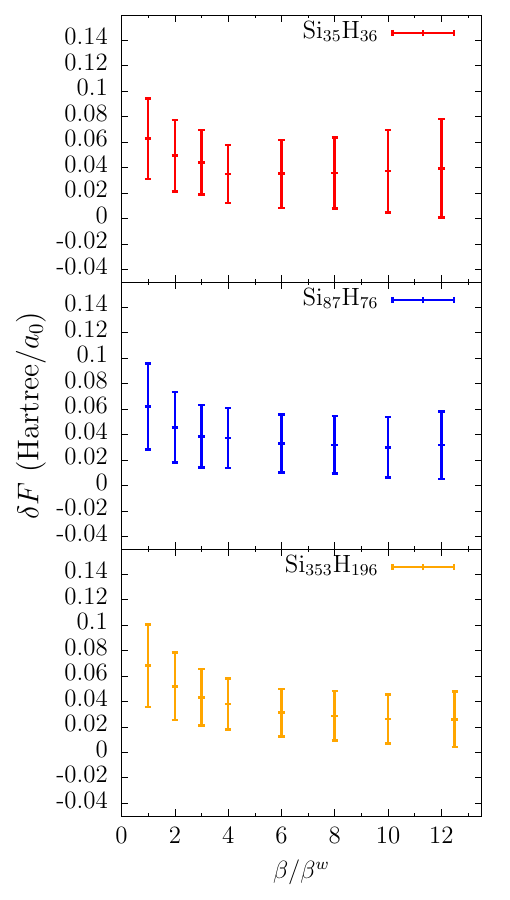}
\par\end{centering}
\caption{\label{fig:Forces} Analogous to Fig. \ref{fig:EperN} but shows $\delta F,$the
error in the averaged force relative to the deterministic value, normalized
over all silicon atoms and over the $N_{{\rm ind}}=10$ runs, and
the associated standard deviation $\sigma_{F}$ (Eqs. (\ref{eq:DeltaF},\ref{eq:SigmaF})).
In sDFT $\delta F$ is significantly larger than $\sigma_{F},$ indicating
some amount of systematic deviation. In t-sDFT, around $\beta/\beta^{w}\sim7-10,$
both the stochastic and especially the systematic errors decrease,
i.e., $\sigma_{F}$ decreases and $\delta F\sim\sigma_{F}.$}
\end{figure}

For each system, we performed calculations for several $\beta/\beta^{w}$
ratios. As these systems are semiconductors, we are simply interested
in the limit where the Fermi-Dirac distribution is effectively a step
function. We, therefore, replace the Fermi-Dirac distribution by the
complementary error function, $f_{\beta\:\mu}(\hat{h})=\frac{1}{2}{\rm erfc}(\beta(\hat{h}-\mu)),$
which looks similar to the Fermi-Dirac distribution but does not require
a very small $\beta$ to be effectively a step function.

The numerical parameters for the runs are summarized in Table~\ref{tab:I}.
There are several points to note. 

First, since the three systems have with increasing size a progressively
smaller band gap, $E_{g}$, the larger systems require larger values
of $\beta$ and correspondingly larger Chebyshev expansion lengths,
$K$. Furthermore, we use (for all systems) $N_{s}=30$ orbitals for
the sDFT calculations. Finally, note that the warm temperature calculations
which require most of the numerical work have an expansion length
$K^{w}$ and a corresponding number of stochastic orbitals $N_{s}^{w}$
chosen so that the total work $W^{tot}$ is the same for each value
of $\beta/\beta^{w}$ (this includes sDFT at $\beta/\beta^{w}=1$).
This allows us to compare the efficacy of tempering in terms of the
reduction of fluctuations as a function of $\beta/\beta^{w}$.

\subsubsection*{Prelude: Stochastic vs. Systematic Deviation}

The results shown below exhibit two kinds of deviations, which are
briefly reviewed; for a fuller discussion see Ref. \citep{fabian2019stochastic}.
The first deviation the usual stochastic Monte-Carlo fluctuation which
scales with the number of samples as $N_{s}^{-1/2}.$ The other kind
is a systematic deviation. Such deviation scales as $N_{s}^{-1}$
and appears whenever the results of the Monte-Carlo sampling are used
in an iterative self-consistent process (see, e.g., \citep{neuhauser2017stochastic}).
Here, since the density is prepared from a finite number of stochastic
orbitals and the filtered stochastic orbitals depend on the density,
the self-consistent SCF procedure has a systematic deviation. 

The practical effect of the systematic deviation is simple to state:
when doing calculations with a finite $N_{s},$ and repeating these
calculations many times, the averaged result would differ from the
true $N_{s}\to\infty$ result. As we show below, in most of our calculations,
tempering reduces and practically eliminates this systematic deviation,
avoiding the need to use jackknife or bootstrap methods \citep{10.2307/2241717}.

\subsubsection*{Energies}

Fig. \ref{fig:EperN} shows, for the three different systems, the
averaged total energies per particle based on the $N_{{\rm ind}}=10$
runs and the associated error bars for different $\beta^{w}$ values.
For simplicity, the results are depicted as a function of $\beta/\beta^{w}$.
In addition, we include the deterministic DFT values for comparison.
Interestingly (see also the SI), for a fixed $N_{s}$, the systematic
deviation decreases by a factor of 2 when the system size increases
by a factor of 10, while the stochastic error decreases by a larger
factor with system size, scaling as $N_{e}^{-1/2}$ due to self-averaging. 

Consider first the starting point for each figure, $\beta/\beta^{w}=1$,
which is simply sDFT (i.e., with no correction terms). Since we only
use $N_{s}=30$ stochastic orbitals, a very small number, the sDFT
calculations show a significant systematic deviation, i.e., the averaged
energy-per-particle is several standard deviations away from the deterministic
value. 

Turning to t-sDFT (i.e., $\beta^{w}<\beta$), we see that both the
systematic deviation and the stochastic error decrease as $\beta/\beta^{w}$
increases. As evident from Fig. \ref{fig:EperN} (and verified by
a second-order polynomial fit of the error in the energy as a function
of $\beta/\beta^{w}$ in the SI) once $\beta/\beta^{w}\sim7-12$ there
is essentially no systematic deviation while the stochastic fluctuations
decrease by a factor of around 4-5.

The reduction in the statistical error and systematic deviation as
$\beta/\beta^{w}$ increases relative to sDFT ($\beta/\beta^{w}=1$)
for a fixed $W^{tot}$ results from the fact that we can use a much
larger number of stochastic orbitals for the warmer temperature density,
the significant contribution to the density, compared to the sDFT
density. Despite the need to use a longer Chebyshev expansion, we
can use fewer stochastic orbitals for the correction term since its
contribution to the density is smaller.

Finally note in Fig. 2, with increasing system size, the optimal $\beta/\beta^{w}$
values (i.e., that results in the smallest small statistical fluctuations
in the energy per electron) shift to larger ratios. This is partially
a result of quantum confinement, so that with increasing size the
KS gap decreases and therefore $\beta$ increases, and partially due
to the modified density of state structure for the larger clusters
which causes the optimal $\beta^{w}$ to increase with system size.
Therefore, the smallest system the optimal ratio is about $\beta/\beta^{w}\sim3-4$,
and for the larger clusters minimum error is obtained for $\beta/\beta^{w}\sim7-12.$

\subsubsection*{Forces and Density}

\begin{figure}
\begin{centering}
\emph{\includegraphics[width=1\columnwidth]{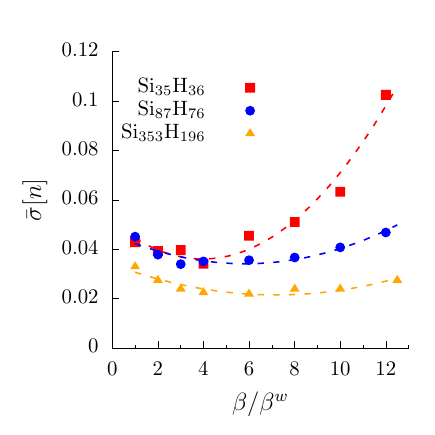}}
\par\end{centering}
\caption{\label{fig:Dens} The normalized integral of the standard deviation
of the density per electron, together with a parabolic fit. For the
smaller system, the density deviation decreases between $\beta/\beta^{w}=2-4$,
and for the two larger systems the stochastic errors decrease around
a larger range $\beta/\beta^{w}=2-10$, by up to 30\%-40\%.}
\end{figure}

We next show how t-sDFT reduces the errors in the atomic forces compared
to sDFT (we only analyze the forces on the silicon atoms for comparison).
Fig. \ref{fig:Forces} is similar to the energy plot in Fig. \ref{fig:EperN},
but here we plot the normalized deviation of the averaged stochastic
forces from the deterministic forces, $\delta F$:

\begin{equation}
(\delta F)^{2}\equiv\frac{1}{N_{{\rm Si}}}\sum_{i=1}^{N_{{\rm Si}}}|\bar{\boldsymbol{F}}^{i}-\boldsymbol{F}^{d,i}|^{2},\label{eq:DeltaF}
\end{equation}
where a bar indicates averaging over the $N_{{\rm ind}}=10$ runs
and ``$d$'' stands for deterministic; $i$ is an index over the
silicon atoms. The error bars in Figure \ref{fig:Forces}, $\sigma_{F},$
indicate the standard deviation of the normalized averaged force of
the silicon atoms, i.e., 
\begin{equation}
\sigma_{F}^{2}=\frac{1}{(N_{{\rm ind}}-1)N_{{\rm ind}}N_{{\rm Si}}}\sum_{j=1}^{N_{{\rm ind}}}\sum_{i=1}^{N_{{\rm Si}}}|\boldsymbol{F}^{i,j}-\bar{\boldsymbol{F}}^{i}|^{2},\label{eq:SigmaF}
\end{equation}
where $\boldsymbol{F}^{i,j}$ is the force over atom $i$ in the $j$'th
independent run.

Note that the magnitude of the errors in the forces is large, but
this could be reduced by increasing the number of independent runs
or stochastic orbitals. However, since the purpose of the study is
to uncover the behavior with respect to $\beta^{w}$, we use a small
number of stochastic orbitals to reduce the computational effort and
thus, apply the approach for many values of $\beta/\beta^{w}$ and
different system sizes. Further, note that since the stochastic errors
are generally not systematic, especially once tempering is used, these
forces can be used for Langevin molecular dynamics if we increase
the number of sampling orbitals by about an order of magnitude. We
have indeed applied sDFT (with an order of magnitude more orbitals
than in this study) in a Langevin molecular dynamics study and have
shown that that the Langevin dissipation matrix is then easily modified
to include the effect of the fluctuations in the sDFT force, and the
correct pair distribution is then obtained \citep{arnon2017equilibrium,Arnon2020langevin}.

As previously discussed, the sDFT forces are similar in the three
systems, since the local environment is similar and therefore the
errors are primarily a function of the number of stochastic orbitals,
$N_{s}$. The reduction in the errors in the forces is appreciable
but less significant than for the energy. Using again a 2nd order
fit (see the SI), we get that the reduction in error in the forces
is about 30\% for the smallest system and goes up to 50\% for the
largest cluster.

To compare the deviation in the density using t-sDFT to sDFT, we use
the integral of the standard deviation of the averaged density per
electron, defined as $\bar{\sigma}[n]\equiv(N_{{\rm ind}}-1)^{-1/2}N_{e}^{-1}\sum\sigma(n(\boldsymbol{r}))dV$,
where $\sigma(n(\boldsymbol{r}$)) is the standard deviation in the
density at grid point $\boldsymbol{r}$. Fig.~\ref{fig:Dens} shows
that tempering again reduces the stochastic error for $\ensuremath{\beta/\beta^{w}}$
values around 7-10. The reduction in the deviation of the density
is similar in magnitude to that of the total atomic forces, up to
30\%-40\%, and is much less dramatic than the error reduction in the
total energies. 

Finally, note that Fig. 4 shows that when the value of $\beta/\beta^{w}$
is very large, the density fluctuations start rising with the $\beta/\beta^{w}$
ratio; for large ratios, the warm density deviates significantly from
the low-temperature density, so the difference between the two densities
is significant which causes large statistical fluctuations.

\section{\label{sec:Summary-and-Discussion} Conclusions}

We presented here a tempering method for stochastic density functional
theory that reduces the statistical error in the total energy. Our
scheme (t-sDFT) relies on decomposing the density into a large high-temperature
term with a correction density. The new method expands the density
in terms of the inverse temperature, $\beta$, to take advantage of
the fact that with lower $\beta$ (i.e., a higher effective temperature)
fewer Chebyshev polynomials are needed, thus enabling the use of more
stochastic orbitals without increasing the computational cost.

A natural extension of this work is the implementation of multiple-$\beta$
tempering with more than two values of $\beta$, as done earlier for
deterministic renormalization-group studies; the formalism is presented
in Appendix A. In this work we have not implemented embedded fragments,
an approach that independently reduces the standard deviation in the
energy and forces. In future work, the two methods will be combined
to hopefully further reduce the stochastic error. Further work will
also explore how to optimize the choice of $\beta$ values and the
number of stochastic orbitals used to reduce the stochastic deviations.

Our method reduces the standard deviation in the total energy by a
factor of around 4-5, which corresponds to reducing the total number
of required stochastic orbitals by more than an order of magnitude.
This is only for the total energies, while the error in the forces
and density is reduced by a smaller amount only 30\%-50\% and 30\%-40\%
respectively. Interestingly, this is the opposite behavior relative
to energy-window sDFT where the error in the forces is improved significantly
but not the error in the total energies. Another interesting aspect
is that sDFT almost eliminates systematic deviation.

The main conclusion of our work is that for the same overall cost,
tempering improves the accuracy by 1.4-4 depending on the quantity
studied while also shrinking the systematic deviation so that the
results are closer to the deterministic value even for a small number
of samples. Equivalently, for the same stochastic deviation, tempering
reduces the overall effort by a factor ranging from $\approx$2 to
$\approx20$, depending on the desired quantity.

\section*{Acknowledgments}

\vspace{-10bp}
This paper was supported by the Center for Computational Study of
Excited State Phenomena in Energy Materials (C2SEPEM), which is funded
by the U.S. Department of Energy, Office of Science, Basic Energy
Sciences, Materials Sciences and Engineering Division via Contract
No. DE-AC02- 05CH11231, as part of the Computational Materials Sciences
Program. In addition, RB gratefully acknowledges the support from
the US-Israel Binational Science Foundation (BSF) under Grant No.
2018368. Computational resources were supplied through the XSEDE allocation
TG-CHE170058.

\section*{Data Availability Statement}

The data that support the findings of this study are available from
the corresponding author upon reasonable request.

\section*{Appendix A: Multiple $\beta$}

The general expansion of the filter $\hat{\rho}_{\beta}$ for $L$
temperatures, ordered so $\beta\equiv\beta_{1}>\beta_{2}>...>\beta_{L}$,
is

\begin{equation}
\hat{\rho}_{\beta}=\hat{\rho}_{\beta_{L}}-\sum_{\ell=1}^{L-1}\Delta\hat{\rho}_{\ell},
\end{equation}
where

\begin{equation}
\Delta\hat{\rho}_{\ell}=\hat{\rho}_{\beta_{\ell+1}}-\hat{\rho}_{\beta_{\ell}}.
\end{equation}
This expansion leads to an expression for the density similar to Eqs.
\ref{eq:warm_dens}-\ref{eq:Deltanr}. The case we studied in the
paper is simply $L=2,$ with $\beta_{1}\equiv\beta$ and $\beta_{2}\equiv\beta^{w}$. 

\bibliographystyle{unsrt}
\bibliography{t-sDFT}

\begin{thebibliography}{10}

\bibitem{jones2015density}
R.~O. Jones.
\newblock Density functional theory: {Its} origins, rise to prominence, and
  future.
\newblock {\em Reviews of Modern Physics}, 87(3):897--923, August 2015.

\bibitem{kolb2012molecular}
Brian Kolb and T~Thonhauser.
\newblock Molecular {Biology} at the {Quantum} {Level}: {Can} {Modern}
  {Density} {Functional} {Theory} {Forge} the {Path}?
\newblock {\em Nano LIFE}, 2(02), 2012.

\bibitem{selli2017modelling}
Daniele Selli, Gianluca Fazio, and Cristiana Di~Valentin.
\newblock Modelling realistic {TiO}$_{2}$ nanospheres: {A} benchmark study of
  {SCC}-{DFTB} against hybrid {DFT}.
\newblock {\em The Journal of Chemical Physics}, 147(16):164701, 2017.
\newblock Publisher: AIP Publishing LLC.

\bibitem{freysoldt2014firstprinciples}
Christoph Freysoldt, Blazej Grabowski, Tilmann Hickel, J{\"o}rg Neugebauer,
  Georg Kresse, Anderson Janotti, and Chris~G. Van~de Walle.
\newblock First-principles calculations for point defects in solids.
\newblock {\em Rev. Mod. Phys.}, 86(1):253--305, March 2014.
\newblock Publisher: American Physical Society.

\bibitem{baer1997sparsity}
Roi Baer and Martin Head-Gordon.
\newblock Sparsity of the {Density} {Matrix} in {Kohn}-{Sham} {Density}
  {Functional} {Theory} and an {Assessment} of {Linear} {System}-{Size}
  {Scaling} {Methods}.
\newblock {\em Phys. Rev. Lett.}, 79(20):3962--3965, November 1997.

\bibitem{scuseria1999linearscaling}
G.~E. Scuseria.
\newblock Linear scaling density functional calculations with {Gaussian}
  orbitals.
\newblock {\em J. Phys. Chem. A}, 103(25):4782--4790, 1999.

\bibitem{ozaki2010efficient}
T.~Ozaki.
\newblock Efficient low-order scaling method for large-scale electronic
  structure calculations with localized basis functions.
\newblock {\em Phys. Rev. B}, 82(7):075131, 2010.

\bibitem{romero-muniz2018highaccuracy}
Carlos Romero-Muniz, Ayako Nakata, Pablo Pou, David~R Bowler, Tsuyoshi
  Miyazaki, and Rub{\'e}n P{\'e}rez.
\newblock High-accuracy large-scale {DFT} calculations using localized orbitals
  in complex electronic systems: {The} case of graphene{\textendash}metal
  interfaces.
\newblock {\em Journal of Physics: Condensed Matter}, 30(50):505901, 2018.
\newblock Publisher: IOP Publishing.

\bibitem{baer2013selfaveraging}
Roi Baer, Daniel Neuhauser, and Eran Rabani.
\newblock Self-{Averaging} {Stochastic} {Kohn}-{Sham} {Density}-{Functional}
  {Theory}.
\newblock {\em Phys. Rev. Lett.}, 111(10):106402, September 2013.

\bibitem{neuhauser2014communication}
Daniel Neuhauser, Roi Baer, and Eran Rabani.
\newblock Communication: {Embedded} fragment stochastic density functional
  theory.
\newblock {\em J. Chem. Phys.}, 141(4):041102, 2014.

\bibitem{chen2019overlapped}
Ming Chen, Roi Baer, Daniel Neuhauser, and Eran Rabani.
\newblock Overlapped embedded fragment stochastic density functional theory for
  covalently-bonded materials.
\newblock {\em J. Chem. Phys.}, 150(3):034106, January 2019.

\bibitem{fabian2019stochastic}
Marcel~D. Fabian, Ben Shpiro, Eran Rabani, Daniel Neuhauser, and Roi Baer.
\newblock Stochastic density functional theory.
\newblock {\em Wiley Interdisciplinary Reviews: Computational Molecular
  Science}, 10.1002/wcms.1412(0):e1412, 2019.

\bibitem{chen2019energywindow}
Ming Chen, Roi Baer, Daniel Neuhauser, and Eran Rabani.
\newblock Energy window stochastic density functional theory.
\newblock {\em J. Chem. Phys.}, 151(11):114116, September 2019.

\bibitem{chen2021stochastic}
Ming Chen, Roi Baer, Daniel Neuhauser, and Eran Rabani.
\newblock Stochastic density functional theory: {Real}- and energy-space
  fragmentation for noise reduction.
\newblock {\em J. Chem. Phys.}, 154(20):204108, May 2021.

\bibitem{baer1998energyrenormalizationgroup}
Roi Baer and Martin Head-Gordon.
\newblock Energy renormalization-group method for electronic structure of large
  systems.
\newblock {\em Physical Review B-Condensed Matter}, 58(23):15296--15299, 1998.

\bibitem{hutchinson1990astochastic}
Michael~F Hutchinson.
\newblock A stochastic estimator of the trace of the influence matrix for
  {Laplacian} smoothing splines.
\newblock {\em Commun Stat Simul Comput.}, 19(2):433--450, 1990.

\bibitem{kosloff1988timedependent}
R.~Kosloff.
\newblock Time-{Dependent} {Quantum}-{Mechanical} {Methods} for {Molecular}-
  {Dynamics}.
\newblock {\em J. Phys. Chem.}, 92(8):2087--2100, 1988.

\bibitem{sankey1994projected}
Otto~F Sankey, David~A Drabold, and Andrew Gibson.
\newblock Projected random vectors and the recursion method in the
  electronic-structure problem.
\newblock {\em Phys. Rev. B}, 50(3):1376, 1994.

\bibitem{perdew1992accurate}
J.P. Perdew and Y.~Wang.
\newblock Accurate and {Simple} {Analytic} {Representation} of the
  {Electron}-{Gas} {Correlation}-{Energy}.
\newblock {\em Phys. Rev. B}, 45(23):13244--13249, 1992.

\bibitem{troullier1991efficient}
N.~Troullier and J.~L. Martins.
\newblock Efficient {Pseudopotentials} for {Plane}-{Wave} {Calculations}.
\newblock {\em Phys. Rev. B}, 43(3):1993--2006, 1991.

\bibitem{kleinman1982efficacious}
Leonard Kleinman and D.~M. Bylander.
\newblock Efficacious {Form} for {Model} {Pseudopotentials}.
\newblock {\em Phys. Rev. Lett.}, 48(20):1425--1428, May 1982.

\bibitem{martyna1999areciprocal}
G.~J. Martyna and M.~E. Tuckerman.
\newblock A reciprocal space based method for treating long range interactions
  in ab initio and force-field-based calculations in clusters.
\newblock {\em J. Chem. Phys.}, 110(6):2810--2821, 1999.

\bibitem{neuhauser2017stochastic}
Daniel Neuhauser, Roi Baer, and Dominika Zgid.
\newblock Stochastic self-consistent second-order {Green}`s function method for
  correlation energies of large electronic systems.
\newblock {\em J. Chem. Theory Comput.}, 13:5396--5403, 2017.

\bibitem{10.2307/2241717}
Jun Shao and C.~F.~J. Wu.
\newblock A general theory for jackknife variance estimation.
\newblock {\em The Annals of Statistics}, 17(3):1176--1197, 1989.

\bibitem{arnon2017equilibrium}
Eitam Arnon, Eran Rabani, Daniel Neuhauser, and Roi Baer.
\newblock Equilibrium configurations of large nanostructures using the embedded
  saturated-fragments stochastic density functional theory.
\newblock {\em J. Chem. Phys.}, 146(22):224111, June 2017.

\bibitem{Arnon2020langevin}
Eitam Arnon, Eran Rabani, Daniel Neuhauser, and Roi Baer.
\newblock Efficient {L}angevin dynamics for
  {\textquotedblleft}noisy{\textquotedblright} forces.
\newblock {\em The Journal of Chemical Physics}, 152(16):161103, April 2020.

\end{thebibliography}

\end{document}